\documentclass[a4paper,pre,twocolumn,amsmath,amssymb,showpacs]{revtex4}

\usepackage{graphicx}

\DeclareMathOperator{\sech}{sech}

\newcommand{\conj}[1]{{#1}^{\ast}}
\newcommand{\order}[1]{{\mathcal O}\left(#1\right)}
\newcommand{\be}{\begin{equation}}
\newcommand{\ee}{\end{equation}}

\begin{document}
\title{Wobbling kinks in $\phi^4$ theory}
\author{I.V. Barashenkov}
\email{igor@odette.mth.uct.ac.za; Igor.Barashenkov@uct.ac.za}
\affiliation{Department of Mathematics, University 
of Cape Town, Rondebosch 7701, South Africa}
\author{O.F. Oxtoby}
\email{Oliver.Oxtoby@gmail.com}
\affiliation{
CSIR Computational Aerodynamics, Building 12, P.O.~Box 395, Pretoria
0001, South Africa
}\date{Last update on \today}

\begin{abstract}
We present a  uniform 
asymptotic expansion of the wobbling kink to any order in the amplitude
of the wobbling mode. The long-range behaviour
of the radiation is described by matching the asymptotic expansions in
the far field and near the core of the kink.
The complex amplitude
of the wobbling mode is shown to obey a simple 
ordinary differential equation with nonlinear damping. 
We confirm the $t^{-1/2}$-decay law 
  for the 
amplitude which was previously obtained on the basis of
energy considerations.
\end{abstract}


\maketitle

\section{Introduction}

Since the early 1960s, the one-dimensional $\phi^4$ theory 
has been among the most heavily utilised models 
of statistical mechanics and
condensed-matter physics \cite{Stats}. At the same time,
it served as a testing ground
for a variety of ideas in topological quantum field theory
\cite{QFT_books} and cosmology \cite{cosmology}.
 The equation of motion for the model reads
\begin{align}
\label{phi4}
\tfrac{1}{2}\phi_{tt}-\tfrac{1}{2}\phi_{xx} - \phi + \phi^3 = 0,
\end{align}
and the fundamental role in  applications is played by
its kink solution, 
 \begin{align}
 \phi(x,t) = \tanh x.
 \label{tanh}
 \end{align}
The $\phi^4$-kinks describe
domain walls  in 
ferromagnets \cite{Ferromagnetic_DW} and ferroelectrics 
\cite{Ferroelectric_DW,Ferroelectric_exc,Aubry}
and represent elementary excitations in the corresponding partition function 
\cite{Ferroelectric_exc}. They were employed to model 
proton transport in hydrogen-bonded chains \cite{hydrogen}
and charge-density waves in polymers and some metals \cite{CDW,RM}. 
Topological defects described by kinks are generated in large numbers
during second-order phase transitions
\cite{phase_transitions}; such transitions
occur in a variety of condensed matter systems 
and are believed to have been made by different parts 
 of the early Universe
\cite{cosmology}.
 In quantum field theory, kinks represent nonperturbative classical 
solutions which undergo subsequent quantisation \cite{QFT_original};
one example concerns ``bags" containing fermions \cite{bags}.
(For more recent quantum physics applications see \cite{QFT_recent}.)

Mathematically, the $\phi^4$ kink has a lot in common with its
sine-Gordon counterpart; the two kinks are the 
simplest examples of topological solitons in one dimension.
There is an important difference though; the sine-Gordon equation
is integrable whereas the $\phi^4$ theory is not. 
Accordingly, the kink-antikink interaction becomes a nontrivial 
matter in the $\phi^4$ case \cite{Aubry,collisions,Getmanov}.
Another (not unrelated) difference is that
unlike the kink of the sine-Gordon equation, the $\phi^4$ kink
 has an internal mode --- an extra degree
of freedom which allows for oscillations in the width of the kink.
Although these oscillations are accompanied by the emission of radiation 
(another manifestation of the nonintegrability of the $\phi^4$ model),
the radiation is quite weak and the oscillations are sustained over
long periods of time. Since the amplitude of the oscillations
can be fairly large, this periodically expanding and contracting kink
(termed {\it wobbling\/} kink in literature, or simply {\it wobbler\/})
can be regarded as one of the fundamental nonlinear excitations of the 
$\phi^4$ theory, on a par with the nonoscillatory kinks
and breathers. For small oscillation amplitudes
and on short time intervals,
the wobbler can be characterised simply as a linear perturbation 
of the stationary kink \eqref{tanh}.
However in order to determine the lifetime
of this particle-like structure (even when its amplitude is small),
or characterise it when it is a large-amplitude excitation,
  one needs a self-consistent 
fully nonlinear description. 

The wobbling kink was discovered in the early numerical experiments of
Getmanov \cite{Getmanov} who interpreted it as a bound state 
of three fundamental (i.e. nonoscillatory) kinks. (For a more recent series
 of numerical simulations,   
see \cite{Belova}.)
 Rice and Mele have reobtained this nonlinear
excitation within
a variational approach employing the width of the kink 
as a dynamical variable \cite{RM,Rice}.
Segur then constructed the 
quiescent (i.e. nonpropagating) 
wobbler
as a regular perturbation expansion in powers of the 
 oscillation amplitude \cite{Segur}. He  calculated 
  the first two orders of the perturbation
series and noted the likely occurrence of unbounded terms at 
the third, $\epsilon^3$-, order, implying the consequent breakdown
of the expansion.
His construction was extended in 
 Ref.\cite{PelSlu} 
 where the effect of the wobbling on the stationary component 
 of the kink was evaluated. It is also appropriate to mention
 Ref.\cite{Roman1} where its author derived 
  an expression for the radiation wave emitted by an initially
  nonradiating wobbler, and a series of publications \cite{radiation} where 
the interaction of the wobbler with radiation waves was studied in
more detail and from a variety
of perspectives. From the fact that the energy of the wobbling kink
is quadratic
in the amplitude of the wobble while the second-harmonic radiation flux 
is quartic, 
it is straightforward to conclude that the amplitude decays 
as $t^{-1/2}$ \cite{Malomed,Manton,Roman1}.

 Moving on to
 singular perturbation expansions, Kiselev \cite{Kiselev_free} 
studied the perturbed 
$\phi^4$ kink using the Krylov-Bogoliubov-Mitropolskii 
method. (Later, he extended his analysis to the 
$\phi^4$ equation with a  conservative 
autonomous perturbation \cite{Kiselev_perturbed}.) 
His two papers have mathematical rigour and a wealth of useful formulas; 
however a
self-consistent system of equations 
for the kink's parameters was not derived in 
\cite{Kiselev_free,Kiselev_perturbed} 
and the long-term evolution of the 
wobbling kink has therefore remained unexplored.
Manton and Merabet \cite{Manton} 
used the Lindstedt-Poincar\'e method \cite{Nayfeh},
where
the expansion of the field is supplemented by
an expansion of the frequency of the wobbling.
 Manton-Merabet's 
theory was successful in reproducing the decay law of the
wobbling amplitude (which was previously
 obtained from the energy considerations \cite{Malomed,Manton,Roman1}). 
However, the Lindstedt-Poincar\'e method, although efficient 
in finding periodic orbits, may
lead to erroneos conclusions about nonperiodic regimes \cite{Nayfeh}. 
(One manifestation of this inadequacy in the case at hand 
is that the nonlinear corrections to the
frequency become complex and time-dependent \cite{Manton};
a less obvious difficulty is the infinite speed of the signal
propagation, see below.) This motivates the search for 
new approaches which would be
mathematically
self-consistent (like the one in \cite{Kiselev_free,Kiselev_perturbed})
on the one hand, and
preserve the physical insights of phenomenological expansions
 \cite{Manton,Roman1} on the other.

The aim of the present paper is to develop a  singular perturbation 
expansion of this kind.  
Our approach  recognises the existence of 
a hierarchy of space and time scales associated with the kink+radiation
system
and generates a perturbative expansion
which remains uniform to all orders. The
consistent treatment of radition requires also the introduction of an 
independent expansion of the far field which is then matched to the 
expansion near the core of the kink. 
 This produces
physically consistent and asymptotically accurate results 
at all space and time scales.
 In particular we will obtain a nonlinear ordinary
differential equation obeyed by the amplitude of the wobbling mode.  
In the follow-up paper \cite{OB} our multiscale approach 
will be used for the analysis of the wobbling kink 
driven by a resonant force.

The basics of our method are outlined in the next section.
In sections \ref{LinCor} and \ref{QuadCor} we evaluate the first- and
second-order corrections to
the shape of the wobbling kink, and in section \ref{AmEq}
derive an equation for the amplitude of the wobbling mode. 
The asymptotic matching of the radiation 
on the short and long scale is carried out in section \ref{long_scale};
here we show, in particular, how to account for finite propagation 
speed of radiation in a mathematically consistent way.
Finally, conclusions of this study are summarised in section \ref{conclusions}.

\section{The method}
\label{Free}

We consider the kink moving with the velocity $v$.
Making the change of variables $(x,t) \to (\xi,\tau)$, where
\[
\xi = x - \int_0^t v(t') dt', \quad \tau=t, 
\]
 we
transform Eq.\eqref{phi4} to the co-moving frame:
\begin{align}
\label{phi4converted}
\tfrac{1}{2}\phi_{\tau \tau} - v\phi_{\xi \tau} - 
\frac{v_\tau}{2} \phi_\xi - \frac{1-v^2}{2}\phi_{\xi\xi} - \phi +
\phi^3 = 0.
\end{align}
Like the authors
of \cite{Sukstanskii}, we shall determine the kink's
velocity $v(\tau)$ by imposing  the condition
that the kink be always centred at $\xi=0$  [i.e. at $x=\int_0^t
v(t') dt'$].

At first glance, the inclusion of the function $v(t)$ is unnecessary:
having constructed a quiescent
wobbling kink, we could make it move at any speed simply 
by a Lorentz boost. The reason we have introduced the velocity
explicitly in Eq.\eqref{phi4converted}, is twofold. Firstly,
this will allow us to check whether the wobbling kink can drift with a 
{\it nonconstant\/} velocity. The soliton moving with a variable $v(t)$ 
could obviously not
be Lorentz-transformed to the rest frame. Secondly,
we include the velocity in preparation for the
analysis of the damped-driven $\phi^4$
equation in the second part of this project \cite{OB}.
Since the damping and driving terms violate relativistic invariance,
the explicit introduction of the velocity becomes essential 
even when considering the damped-driven wobblers 
moving at a constant speed.

We expand the field about the kink
$\phi_0 \equiv \tanh \xi$:
\begin{align}
\label{phiexpans}
\phi &= \phi_0 + \epsilon\phi_1 + \epsilon^2\phi_2 + \ldots.
\end{align}
Here $\epsilon$ is a (formal) small parameter; it will drop out of
the final expression for the solution [see Eq.\eqref{phi_sum} below]. 
Substituting  \eqref{phiexpans} in \eqref{phi4converted}
and setting to zero coefficients of like powers of $\epsilon$
would constitute Segur's approach which is expected to produce 
secular terms in the expansion \cite{Segur}. To avoid these,
we  introduce a sequence of stretched space
and time variables
\begin{equation}
X_n \equiv \epsilon^n \xi, \quad
T_n \equiv \epsilon^n \tau, \quad n=0,1,2,...,
\end{equation}
which describe slower times and longer distances.
In the limit $\epsilon \to 0$,
the different scales become uncoupled and may be treated as
independent variables. 
We expand the $\xi$- and $t$-derivatives 
 in 
terms of the scaled variables by using the chain rule, 
\begin{align}
\frac{\partial}{\partial \xi} &= 
\partial_0 + \epsilon \partial_1 
+ \epsilon^2 \partial_2 + \ldots, \nonumber  \\
\frac{\partial}{\partial \tau} &= D_0 + \epsilon D_1 
+ \epsilon^2 D_2 + \ldots, \label{chain}
\end{align}
where we have used the standard short-hand notation
\[
\partial_n   \equiv   \frac{\partial}{\partial X_n}, \quad
D_n   \equiv \frac{\partial}{\partial T_n}.
\]
Substituting these expansions into the $\phi^4$ equation 
\eqref{phi4converted}, along with the series
\eqref{phiexpans}, and equating coefficients of like powers of $\epsilon$,
we obtain a hierarchy of equations.
We assume that the velocity of the kink is slowly varying and, for simplicity, that it
is small, i.e. $v = \epsilon V$  where
$V = V(T_1, T_2, \ldots)$ is of order 1.

\section{Linear perturbations}
\label{LinCor}

At $\order{\epsilon^1}$, we obtain the linearisation of Eq.\eqref{phi4}
about the kink $\phi_0=\tanh X_0$:
\be \label{linear}
\tfrac{1}{2}D_0^2\phi_1 + {\mathcal L}\phi_1 = 0,
\ee
where we have introduced the Schr\"odinger operator
\begin{equation}
{\mathcal L} = -\tfrac{1}{2}\partial_0^2-1+3\phi_0^2 = 
-\tfrac{1}{2}\partial_0^2+2-3\sech^2X_0.
\label{calL}
\end{equation}

The general solution of the variable-coefficient
 Klein-Gordon equation 
 \eqref{linear}
can
be written as
\be
\phi_1= C y_T(X_0)
+ A e^{i \omega_0 T_0} y_w(X_0) + c.c. + \phi_R(X_0,T_0),
\label{lin_cor} 
\ee
with
\be 
\phi_R= 
\int_{-\infty}^{\infty} 
\left[ {\cal R} (p)e^{i \omega(p) T_0}+ {\cal R}^*(-p)e^{-i \omega(p) T_0}
\right]
 y_p(X_0) dp. 
 \label{int_rad}
 \ee
 Here $y_T$ and $y_w$ are eigenfunctions of the operator $\cal L$
associated with eigenvalues $0$ and $\frac32$, respectively:
\begin{eqnarray}
y_T(X_0)= \sech^2X_0; \label{y0} \\
y_w(X_0)= \sech X_0 \tanh X_0. \label{y1} 
\end{eqnarray}
The eigenfunction $y_w$ gives the spatial 
profile of the so-called internal mode, also
known as the {\it wobbling mode\/} in the current context.
The internal mode frequency $\omega_0= \sqrt{3}$.
The functions $y_p(X_0)$ are solutions pertaining to
the continuous spectrum of $\cal L$:
\be
{\cal L} y_p = \left( 2+ \frac{p^2}{2} \right) y_p,
\quad -\infty<p<\infty;
\ee
 these were constructed by Segur \cite{Segur}:
 \begin{subequations}
\label{segurssolutions}
\begin{multline}
y_p(X_0) = e^{ipX_0}\Bigg[1+\frac{3(1-ip)}{1+p^2}\tanh X_0 (1+\tanh X_0) \\
- \frac{3(2-ip)}{4+p^2}(1+\tanh X_0)^2\Bigg].
\end{multline}
We also mention an equivalent form for
these solutions:
\begin{multline}
y_p(X_0) = \frac{1}{(1+ip)(2+ip)}e^{ipX_0}\\
\times (2-p^2-3ip \tanh X_0 - 3 \sech^2 X_0).
\end{multline}
\end{subequations}
The continuous spectrum solutions are usually
referred to as phonon modes; 
the phonon frequencies $\omega(p)$ are given by
$\omega(p)=\sqrt{4+p^2}>0$.  
Finally, the coefficients ${\cal R}(p)$ and $A$ are complex; $C_0$ is real,
and $c.c.$ in \eqref{lin_cor}
 stands for the complex conjugate of the immediately
preceding term.

The $y_T$ eigenfunction is the translation mode; 
since the kink is assumed to be centered at $X_0=0$, we let
$C=0$. Next, we will consider only localised
perturbations of the kink, i.e. we assume that $\phi_1 \to
 0 $ as $|X_0| \to \infty$.
 This means that the Fourier coefficient ${\cal R}(p)$ in
 the integral  \eqref{int_rad} can be 
 regarded as an ordinary function, i.e. it does not include any
 $\delta$-function contributions. 
 Sending $T_0 \to \infty$  for the fixed finite $X_0$, 
 Kelvin's formula of the method
 of stationary phase gives 
 \[
 \phi_R(X_0,T_0) \to
 \left( \frac{4 \pi}{T_0} \right)^{1/2}
 {\cal R}(0) y_{0}(X_0) e^{2i T_0+ i \pi/4}  + c.c.
 \]
  Therefore $\phi_R$ is a slowly-decaying wavepacket which will be
 dominated by the zero-wavenumber radiation after other harmonics
 have dispersed away. 
 
 Since we are interested  in the evolution of the 
 wobbling mode and not that of a general localised initial condition, 
 we  set ${\cal R}(p)=0$.    
 Therefore the first-order perturbation is taken in the form
\begin{equation}
\phi_1 =
 A(X_1, \ldots; T_1, \ldots)
 \sech X_0\tanh X_0e^{i \omega_0 T_0} + c.c.
 \label{phi_1}
\end{equation}
The amplitude of the wobbling mode, $A$, is
constant with respect to  $X_0$ and $T_0$ but may depend on slower
times and longer distances.

\section{Quadratic corrections} 
\label{QuadCor}

At the second order in the perturbation expansion we arrive
at a nonhomogeneous variable-coefficient Klein-Gordon
equation 
\be
\tfrac{1}{2}D_0^2\phi_2 + {\mathcal L}\phi_2 
= F_2(X_0,...; T_0,...),
\label{qua}
\ee
where the forcing term is
\begin{subequations} \label{F2}
\begin{eqnarray}
F_2= (\partial_0\partial_1 - D_0 D_1)\phi_1 
- 3\phi_0\phi_1^2+ VD_0\partial_0\phi_1 \nonumber \\
+ \frac{1}{2}D_1V\partial_0\phi_0-\frac{1}{2}V^2\partial_0^2\phi_0.
\label{F2a}
\end{eqnarray}
Substituting for $\phi_0$ and $\phi_1$, this becomes
\begin{eqnarray}
 F_2 = -6|A|^2\sech^2X_0\tanh^3X_0 + \frac{1}{2}D_1V \sech^2X_0
 \nonumber \\
+ V^2 \sech^2X_0 \tanh X_0 
 + \left[ \partial_1A (2\sech^3X_0 \right. \nonumber \\ 
 \left. -\sech X_0) 
- i \omega_0 D_1A\sech X_0\tanh X_0\right. \nonumber \\
 \left. + i\omega_0 VA(2\sech^3X_0-\sech X_0)\right]e^{i \omega_0 T_0} + c.c.
 \nonumber  \\
 - 3A^2\sech^2 X_0\tanh^3X_0e^{2i \omega_0 T_0} + c.c. \ \
\label{eps2eqn} 
\end{eqnarray}
\end{subequations}
The $T_0$-independent term in Eq.\eqref{eps2eqn}
and the term proportional to $e^{i\omega_0 T_0}$
are resonant with the two discrete eigenmodes
of the operator in the left-hand side of \eqref{qua}, while
the term proportional to $e^{2i\omega_0 T_0}$
is resonant with its continuous spectrum.  
The latter part of the forcing is localised in the region near
the origin and acts as a source of radiation
 which spreads outward from there.

We discard the homogeneous solution of Eq.\eqref{qua}
for the same reason as we have discarded most terms in 
the solution of Eq.\eqref{linear}; namely, we do not
want the evolution of the wobbling
mode to be shaded by dispersive transients.
Hence the solution that is of interest to us will consist only of 
the harmonics present in Eq.\eqref{F2}:
\begin{align}
\phi_2 = \varphi_2^{(0)} + \varphi_2^{(1)}e^{i\omega_0 T_0} + c.c. 
+ \varphi_2^{(2)}e^{2i\omega_0 T_0} + c.c.,
\label{phi2m}
\end{align}
where $\varphi_2^{(0)}$, $\varphi_2^{(1)}$ and
$\varphi_2^{(2)}$  are functions of 
$X_0$ which satisfy the three linear 
nonhomogeneous equations:
\begin{eqnarray}
{\mathcal L}\varphi_2^{(0)} = -6|A|^2\sech^2X_0\tanh^3X_0 \nonumber \\
+ \frac{1}{2}D_1V\sech^2X_0 + V^2\sech^2X_0\tanh X_0,
 \label{ph20} \\
({\mathcal L}-\tfrac{3}{2})\varphi_2^{(1)} =  \; 
\partial_1A(2\sech^3X_0-\sech X_0)\nonumber \\
- i \omega_0 D_1A\sech X_0\tanh X_0 \nonumber \\
 + i\omega_0 VA(2\sech^3X_0-\sech X_0), 
 \label{ph21} 
 \end{eqnarray}
 and
 \be
\label{linearspreading}
({\mathcal L}-6)\varphi_2^{(2)} = - 3A^2\sech^2X_0\tanh^3X_0.
\ee
(The functions $\varphi_2^{(0)}$, $\varphi_2^{(1)}$ and
$\varphi_2^{(2)}$ can also depend, parametrically, on $X_1, X_2, ...$
and $T_1, T_2, ...$.)

 The homogeneous solutions of the
first two of these equations are given by 
the eigenfunctions of the operator
$\cal L$, Eqs.\eqref{y0} and \eqref{y1}. 
According to the Fredholm alternative,  the nonhomogeneous equations
admit bounded solutions if and only if
their right-hand sides are orthogonal to the 
corresponding homogeneous solutions. 
For this to be the case, we must set $D_1V = 0$ and $D_1A = 0$.
The variation of parameters yields then
\be
\begin{split}
\varphi_2^{(0)} = 2 |A|^2 \sech^2X_0 \tanh X_0 + \\
+ \left( \frac{V^2}{2}-3|A|^2 \right)
X_0\sech^2X_0 
\label{varphi2}
\end{split}
\ee
and
\be
\label{phi2firstharm}
\varphi_2^{(1)} = -(\partial_1A + i\omega_0 VA)X_0\sech X_0\tanh X_0.
\ee

Although the function $\varphi_2^{(1)}$ decays
to zero as $|X_0| \to \infty$, the product
 $\epsilon \varphi_2^{(1)}$ becomes greater than 
the first-order perturbation $y_1(X_0)$ for each fixed $\epsilon$
and sufficiently large $|X_0|$.
Consequently, the term $\epsilon^2 \phi_2$ in the expansion \eqref{phiexpans}
becomes greater than the previous term, $\epsilon^1 \phi_1$,
leading to nonuniformity of the expansion.
In order to obtain a uniform expansion, we 
 set this ``quasisecular" term to zero:
\be
\partial_1 A + i\omega_0 VA = 0,
\ee
whence 
\be
A=  {\tilde A} (X_2,X_3,...;T_2,T_3,...)e^{-i \omega_0 V X_1}.
\label{A_tilde}
\ee

We also note the terms proportional to 
$X_0 \sech^2 X_0$ in Eq.\eqref{varphi2}. These terms
do not grow bigger than the previous term, $\phi_0= \tanh X_0$, yet
they become larger than the difference 
$\phi_0-1$ as $X_0 \to \infty$ and $\phi_0+1$ as $X_0 \to -\infty$. 
If we attempted to construct the asymptotic expansion 
of the function $\phi-1$ at the right infinity 
or the function $\phi+1$ at the left infinity, the terms
in question would cause nonuniformity of these expansions.
Since the function $X_0 \sech^2 X_0$
is nothing but the derivative of $\tanh (k X_0)$ with respect to $k$,
these terms represent the variation of the kink's
 width. Hence the potential nonuniformity of
 the expansion can be avoided simply by incorporating 
 them in the variable width [see Eq.\eqref{phi_sum} below].
 
We now turn  to  the remaining nonhomogeneous equation,
Eq.\eqref{linearspreading}.
The variation of parameters gives
\be
\label{wrongsoln}
\varphi_2^{(2)} = A^2f_1(X_0),
\ee
with
\begin{multline}
f_1(X_0) = \tfrac{1}{8} \big\{ 6\tanh X_0 \sech^2 X_0 \\
+ (2+ik_0 \tanh X_0+ \sech^2 X_0) [\conj{J}_2(X_0)-J_2^{\infty}] e^{ik_0 X_0} \\
+ (2 -ik_0 \tanh X_0+ \sech^2 X_0)J_2(X_0) e^{-ik_0X_0} \big\}.
\label{f1}
\end{multline}
Here the function $J_2(X_0)$ is defined by the integral
\be
\label{Jdef}
J_2(X_0) = \int_{-\infty}^{X_0} e^{ik_0\xi} \sech^2 \xi \; d\xi 
\ee
with  $k_0=\sqrt{8}$. The constant  $J_2^{\infty}$ is the
asymptotic value of $J_2(X_0)$ as $X_0 \to \infty$:
\be
J_2^{\infty}  =\lim_{X_0 \rightarrow \infty}
J_2(X_0).
\ee

The two constants of integration were chosen such that
the solution \eqref{wrongsoln}-\eqref{f1}
 describes right-moving radiation for positive 
$X_0$ and left-moving radiation for negative $X_0$.
It is not difficult to show  that $f_1$ is an odd function;
we will use this fact in what follows.

\section{Radiation in the far field}
\label{long_scale}

The function \eqref{wrongsoln} is bounded 
but does not decay to zero as $|X_0| \to \infty$. This fact presents a
problem, both for the consistency of our method
 and for
the physical interpretation of the resulting solution.
Mathematically, the term $\epsilon^2 \phi_2$ 
turns out  to be  greater than the previous term
in the expansion \eqref{phiexpans} for sufficiently large $|X_0|$.
As we have mentioned in connection with the term $\varphi_2^{(1)}$,
this leads to nonuniformity of the expansion.
Physically, the problem is that 
any variation of the amplitude of the 
wobbling mode, $A$, on the time scale $T_2$, will result in a
simultaneous change in the amplitude of the radiation tail for all
values of $X_0$, from the origin to the plus- and minus-infinity.
This is obviously in contradiction with the finiteness of the 
velocity of signal propagation in a relativistic
theory [which is bounded by $1$ in the dimensionless units of Eq.\eqref{phi4}.]

The problem stems from the fact that
the  equation \eqref{qua}
and, therefore, equation \eqref{linearspreading}, were obtained 
under the assumption that, in the expansion \eqref{phiexpans},
the second term is smaller than the first one, the third one is
smaller than the second, and so on --- more precisely,
that $\epsilon \phi_1/\phi_0 \to 0$, $\epsilon^2 \phi_2/(\epsilon
\phi_1) \to 0$, and so on, as $\epsilon \to 0$.
This assumption turns out to be only valid 
on the short scale  and therefore,
the equation \eqref{linearspreading} is only
meant to hold for distances $X_0={\cal O}(1)$ but not 
$X_0={\cal O}(\epsilon^{-1})$ or longer.
The interval of $X_0$ where $\epsilon \phi_{n+1}/\phi_n \to 0$ as
$\epsilon \to 0$ will be referred to as 
the ``inner" region in what follows.
Eqs.\eqref{qua}
and \eqref{linearspreading} are therefore valid in the inner region.

To obtain a uniform expansion on the whole axis, we also
consider two ``outer" regions ---  one  with $X_0>0$
and the other one with $X_0<0$. 
We define the outer regions
by the requirement that $|X_0|$ be greater than $\frac{1}{2}
\ln \epsilon^{-1}$.
Note that the outer regions overlap with the
inner region.
For example, the values $X_0= \pm \frac{2}{3} \ln\epsilon^{-1}$
are clearly in the outer regions; on the other hand,
 we have
$\epsilon \phi_1/\phi_0 \to 0$, $\epsilon^2 \phi_2/(\epsilon
\phi_1) \to 0$, {\it etc.\/}
 for these $X_0$ and so
they belong to the inner region as well.
 
In the right outer region, we expand $\phi$ in the power series
 \begin{subequations}
 \label{outexp}
 \begin{equation}
 \phi=1+ \epsilon^2 \phi_2+ \epsilon^4
\phi_4+\dots ,
\end{equation}  and in the left outer region, we let 
\begin{equation}
\phi=-1+ \epsilon^2 \phi_2+ \epsilon^4
\phi_4+\dots . 
\end{equation}
\end{subequations}
Substituting these, together with the expansions 
\eqref{chain},
 in Eq.\eqref{phi4converted}, 
 the order $\epsilon^2$ gives
  \[ \tfrac12 D_0^2 \phi_2 + {\cal L} \phi_2=0,\]
where ${\cal L}=-\frac12 \partial_0^2+2$ is the 
far-field asymptotic
form of the operator \eqref{calL}. The solutions of this equation
in the right and left outer regions are, respectively,
 \begin{subequations}
 \label{outer}
 \begin{equation}
 \phi_2= {\cal J} B_+ e^{i(\omega_+ T_0-k_+ X_0)}+c.c.
 \end{equation} 
 and 
 \begin{equation}
 \phi_2= -{\cal J} B_- e^{i(\omega_- T_0-k_- X_0)}+c.c.,
 \end{equation}
 \end{subequations} 
where 
$\omega_{\pm}^2= k_{\pm}^2+4$,
and the amplitudes $B_\pm$ are functions
of the ``slow" variables: $B_\pm=B_\pm(X_1,...; T_1,...)$. 
The normalisation constant ${\cal J}$ will be chosen at a 
later stage, and the
negative sign in front of $B_-$ 
is also introduced for later convenience.

Eqs.\eqref{outer} should be matched to the solution 
in the inner region, Eq.\eqref{phi2m} with
coefficients as in  \eqref{varphi2},
\eqref{phi2firstharm}, and \eqref{wrongsoln}.
To this end,
we take the values $X_0= \pm \tfrac 23 \ln \epsilon^{-1}$
(which, as we remember, belong to the overlap regions). 
For these $X_0$, we have 
 $|X_1|=\order{\epsilon \ln \epsilon^{-1}}$, $|X_2|=\order{\epsilon^2 \ln
\epsilon^{-1}}$,...,
and so
$X_1 \to 0$, $X_2 \to 0$,..., as $\epsilon \to 0$.
The solutions 
\eqref{outer} become, in this limit:
\begin{equation*}
\phi_2= \pm {\cal J} B_{\pm} (0,0,...; T_1, T_2, ...)
e^{i(\omega_{\pm}T_0-k_{\pm} X_0)} + c.c.
\end{equation*}
On the other hand, letting $|X_0|=\frac23 \ln \epsilon^{-1}$ 
and sending $\epsilon \to 0$ in 
Eqs.\eqref{varphi2},
\eqref{phi2firstharm}, and \eqref{wrongsoln}, we get
\[
\phi_2= \pm (2-i k_0) J_2^\infty A^2(0,0, ...; T_2, T_3, ...)
e^{i(2 \omega_0T_0 \mp k_0 X_0)} + c.c.,
\]
where the top and bottom sign pertain to the 
positive and negative $X_0$, respectively.
Choosing ${\cal J}=(2-i k_0) J_2^\infty$ and
equating the above two expressions, we obtain 
$\omega_{\pm}=2\omega_0$, $k_\pm=\pm k_0$, and
\begin{equation}
B_\pm(0,0,...;T_1,T_2,...)= A^2(0,0,...;T_2,T_3,...).
\label{match} 
\end{equation}

Eqs.\eqref{match} can be regarded as
 the boundary conditions for the amplitude fields
$B_+$ and $B_-$. Equations governing the evolution of
these functions of slow variables can be derived at higher orders 
of the (outer) perturbation expansion. 
Namely, the solvability
condition  at the order $\epsilon^3$ yields
\begin{equation}
(\partial_0 \partial_1-D_0 D_1 +V \partial_0 D_0)\phi_2=0.
\label{swas}
\end{equation}
Substituting from \eqref{outer}, this becomes
\begin{align}
D_1B_{\pm} + \frac{k_{\pm}}{2\omega_0} \partial_1B_{\pm} + i
k_{\pm} V B_{\pm}=0,
\label{3terms}
\end{align}
whence 
\begin{equation}
B_\pm=e^{-2i\omega_0 V X_1} {\cal B}_\pm(X_1,X_2,...; T_1,T_2,...),
\end{equation}
where ${\cal B}_\pm$ satisfy a pair of linear transport equations
\begin{subequations} \begin{eqnarray}
D_1{\cal B}_+ +  c_0 \partial_1{\cal B}_+ =0, \quad X_1 >0, 
\label{transport1}\\
D_1{\cal B}_- -  c_0 \partial_1{\cal B}_- =0, \quad X_1 <0,
\label{transport2}
\end{eqnarray}
\label{transport}
\end{subequations}with $c_0=k_0/(2 \omega_0)$. 
Note that $c_0$ is nothing but the
group velocity of the radiation waves with the wavenumber $k_0$:
$c_0=\left. (d \omega/dk) \right|_{k=k_0}$, where
$\omega=\sqrt{4+k^2}$.

Solution of equations \eqref{transport} with the 
boundary condition \eqref{match} is a textbook exercise.
Assume that the functions ${\cal B}_\pm$  satisfy
the initial conditions
${\cal B}_+(X_1,0)= {\cal B}^{(0)}(X_1)$ (for $X_1>0$) and 
${\cal B}_-(X_1,0)= {\cal B}^{(0)}(X_1)$ (for $X_1<0$), with some
 function ${\cal B}^{(0)}(X_1)$  defined on the whole axis
$-\infty < X_1 < \infty$, with ${\cal B}^{(0)}(X_1) \to 0$
as $|X_1| \to \infty$.
(We have suppressed the 
dependence on the variables $X_2,X_3,...; T_2,T_3,...$ for
notational convenience.) In the region $X_1>c_0 T_1$, the solution to
the equation \eqref{transport1} with the above initial 
condition is given by
${\cal B}_+(X_1,T_1)={\cal B}^{(0)}(X_1 -c_0 T_1)$. 
This solution represents an envelope of a group of 
second-harmonic radiation waves, moving to
the right with the velocity $c_0$. 
Importantly, the amplitude ${\cal B}_+$ in this region
is not related to the wobbling amplitude $A$ and so no information
from the core of the kink can reach
this region.
In the region $0<X_1<c_0 T_1$, the solution 
to Eq.\eqref{transport1} is determined by the boundary condition
instead:
${\cal B}_+(X_1,T_1)=A^2(0;0)$. This result implies that 
the moving envelope has the form of a propagating 
front, leaving 
${\cal B}_+$  flat and stationary in its wake.
In a similar way,
on the negative semiaxis we have a front moving  
with the velocity $-c_0$ and leaving ${\cal B}_-(X_1,T_1)$
equal to the constant $A^2(0;0)$ in its wake.
 
The above analysis has two shortcomings. One drawback is that
we have restricted ourselves to groups of radiation
waves with the characteristic length and time scale 
of order $\epsilon^{-1}$.
A natural question therefore is whether
 variations with larger space and time scales 
 (e.g. variations on $X_2$ and $T_2$ scales) could not propagate
faster than $c_0$. Another latent defect is that the solutions
for ${\cal B}_\pm(X_1,T_1)$ that we have constructed, will generally be
discontinuous along the lines $X_1= \pm c_0 T_1$.
To address both of these issues, we proceeed to the order $\epsilon^4$ 
of the outer expansion where the solvability condition for the
second harmonic gives
\begin{align*}
i(2 \omega_0 D_2+ k_0 \partial_2) B_\pm 
+\frac12 (D_1^2-\partial_1^2) B_\pm \\
+iV (k_0D_1-2 \omega_0 \partial_1) B_\pm
-\frac12 V^2 k_0^2 B_\pm=0.
\end{align*}
Eliminating $D_1 B_\pm$ using \eqref{3terms}, this becomes
\begin{align}
iD_2 B_\pm \pm i c_0 \partial_2 B_\pm- iV \partial_1 B_\pm
-
\frac{\omega_{kk}}{2} \partial_1^2 B_\pm=0,
\label{2nd_cor} 
\end{align}
where $\omega_{kk} \equiv \left. (d^2\omega/dk^2)\right|_{k_0}=
(4 \omega_0^2-k_0^2)/(8 \omega_0^3)$ is the dispersion of
the group velocity of the radiation waves. Combining 
Eq.\eqref{2nd_cor} with \eqref{3terms}, we obtain a pair of
equations in the original space and time variables:
\begin{align}
i \partial_t  B_\pm  \pm ic_0  \partial_x  B_\pm
 \mp vk_0  B_\pm -\frac{\omega_{kk}}{2} \partial_x^2 B_\pm=0. 
\label{real_coo} 
\end{align}

The  pair of linear Schr\"odinger equations \eqref{real_coo}
govern the 
evolution of the radiation amplitudes over times
and distances as large as $\epsilon^{-2}$; if we want to 
have a description on even a larger scale, we
simply need to include equations from higher
orders of the outer expansion.
 Solutions of Eqs.\eqref{real_coo} with
 the boundary conditions $B_\pm=A^2$ at $x=vt$ and 
$B_\pm=0$ at $x= \pm \infty$
have the form of slowly dispersing fronts propagating at
the velocities $\pm c_0$ and interpolating,
{\it continuously}, between $A^2$ and $0$. As in our
previous description exploiting the transport equations \eqref{transport}
and valid on a shorter space-time scale, perturbations of
$A^2$ cannot travel faster than $c_0$, the group
velocity of radiation.

Thus, by 
introducing 
 the long-range variables $B_\pm$,  
 ``untied"  from the short-range amplitude
 $A$,
we have restored the finiteness of the velocity of the 
radiation wave propagation. By introducing the outer expansions,
we have also prevented the breakdown of the asymptotic
expansion at large distances.

\section{Decay law for the wobbling amplitude} 
\label{AmEq}

Returning to the original, ``inner", expansion 
\eqref{phiexpans} and
collecting terms of order $\epsilon^3$ gives the equation
\begin{subequations}
\begin{equation}
\tfrac{1}{2}D_0^2\phi_3 + {\mathcal L}\phi_3 =F_3,
\end{equation}
where
\begin{multline}
\label{epscubed}
F_3 = (\partial_0\partial_1 - D_0 D_1)\phi_2 + (\partial_0\partial_2 - D_0 D_2)\phi_1 \\
+ \tfrac{1}{2}(\partial_1^2-D_1^2)\phi_1 -\phi_1^3 - 6\phi_0\phi_1\phi_2 + VD_0\partial_0\phi_2 \\
+VD_0\partial_1\phi_1 
+ VD_1\partial_0\phi_1 
+ \tfrac{1}{2}D_2V\partial_0\phi_0 -\tfrac{1}{2}V^2\partial_0^2\phi_1.
\end{multline}
\end{subequations}
Having evaluated $F_3$ using the known functions $\phi_0$,
$\phi_1$ and $\phi_2$, we decompose the solution $\phi_3$ into 
 simple harmonics as we did at $\order{\epsilon^2}$.
The solvability condition for
the zeroth harmonic in equation \eqref{epscubed}
gives $D_2V = 0$, which means that $V$ remains constant
up to times $t \sim \epsilon^{-3}$. 
The solvability condition for 
the first harmonic produces
\begin{align}
\label{freeamp}
 i \frac{2  \omega_0 }{3} D_2A + \zeta |A|^2A - V^2A = 0,
\end{align}
where
\begin{eqnarray}
\zeta =6
\int_{-\infty}^{\infty} \sech^2X_0\tanh^3X_0\Big[\tfrac{5}{2}\sech^2X_0\tanh
X_0 \nonumber \\
 -3X_0\sech^2X_0 + f_1(X_0)\Big]\,dX_0. \  \
\end{eqnarray}
Out of the real and imaginary part of $\zeta$, the imaginary
part is more important; 
it can be easily evaluated analytically:
\be
\zeta_I = \frac{3\pi^2k_0}{\sinh^2 \left(\pi k_0/2 \right)}
=  0.04636.
\ee
The real part was computed numerically:
\be
\zeta_R = -0.8509. \label{Rek1}
\ee

Denoting $\epsilon {\tilde A} \equiv a$ the ``natural" (unscaled)  amplitude of 
the wobbling mode, and recalling that $v = \epsilon V$
and $A_t= \epsilon^2 D_2 A +
\order{\epsilon^3}$, we express the amplitude equation
\eqref{freeamp} in terms of the original variables:
\be
\label{mainampeqfree}
i a_t = -\frac{\omega_0 \zeta}{2}   \, |a|^2a + \frac{\omega_0}{2} 
 v^2a +
\order{|a|^5}.
\ee
Eq.\eqref{mainampeqfree} 
contains solvability conditions at all orders covered so far ---
they arise simply by expanding 
the derivative $d/d t$ as in Eq.\eqref{chain}.
Unlike the amplitude equation $D_1A=0$ which only
governs the evolution for times $t \sim \epsilon^{-1}$, 
and unlike the equation \eqref{freeamp} which only
holds on the timescale $ t \sim \epsilon^{-2}$,
the ``master equation" \eqref{mainampeqfree}
is applicable for all times, from $t=0$ to $t \sim \epsilon^{-2}$.

 The master equation \eqref{mainampeqfree}
is
the final result of the asymptotic analysis.
All the conclusions about the behaviour of the 
wobbler's amplitude shall be made on the basis 
of this equation.
We could extend the range of applicability of the 
master equation beyond times of order $\epsilon^{-2}$ by continuing our 
perturbation analysis
to higher orders of $\epsilon$. However, corrections to the
equation \eqref{mainampeqfree} obtained in this way
would be smaller than the terms that are already in the right-hand
side of \eqref{mainampeqfree} and would not affect our conclusions
based on  \eqref{mainampeqfree} in its present form.

The absolute value of $a$ is governed by the equation
\begin{equation} 
\label{decay_law}
\frac{d}{d t}|a|^2 = -\omega_0 \zeta_I \, |a|^4 +
\order{|a|^6}.
\end{equation}
Previously this equation was obtained using heuristic
considerations \cite{Malomed,Manton,Roman1}. 
Since $\zeta_I>0$, the amplitude of
the wobbling is monotonically
decreasing with time: a constant emission of radiation damps the wobbler.
Dropping
the $\order{|a|^6}$ correction term from \eqref{decay_law},  
the decay law is straightforward:
\begin{equation}
 \label{decay_rate}
|a(t)|^2 = \frac{|a(0)|^2} 
{1+ \omega_0 \zeta_I \, |a(0)|^2t }=
\frac{|a(0)|^2} 
{1+ 0.08030 \times |a(0)|^2t }.
\end{equation} 
When $a(0)$ is small, the decay becomes 
appreciable only after long times  $t \sim |a(0)|^{-2}$.  
The decay is slow; for times $t \gg 12.5 \times |a(0)|^{-2}$, 
Eq.\eqref{decay_rate} gives $|a| \sim t^{-1/2}$.

\begin{figure}
  \includegraphics[height = 2.0in,width = 0.95\linewidth]{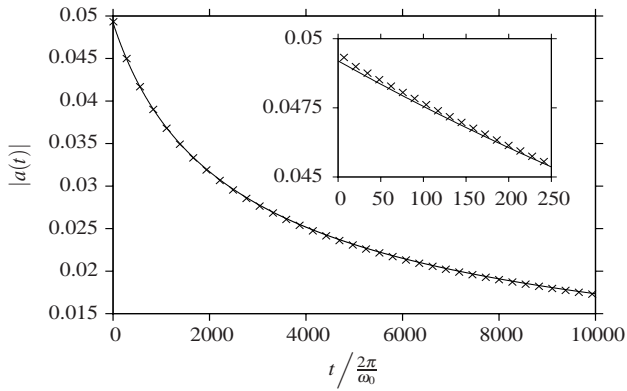}
  \caption{The decay of the free wobbling kink.  Crosses: 
    $|a(t)|$ as obtained from the direct numerical simulations
  of Eq.\eqref{phi4} with the initial conditions in the form
  $\phi=\tanh x +2a_0 \sech x \tanh x$,
  $\phi_t=0$, with $a_0=0.05$. Solid curve: equation 
  \eqref{decay_rate} with $|a(0)|=0.0492$.
  The inset shows the first 250 periods of oscillation;
  during this time the amplitude drops by less than $10\%$.}
  \label{decay_figure}
  \end{figure}

We have verified the above decay law in 
direct numerical simulations of the full partial differential 
equation \eqref{phi4}.
(The details of our numerical algorithm have been relegated to
the Appendix.)
 As the
 initial conditions, we took $\phi(x,0)=\tanh x +2a_0 \sech x \tanh x$
with some real $a_0$ and
  $\phi_t(x,0)=0$. After a short initial transient,
  the solution was seen to settle to the curve  
  \eqref{decay_rate} with $|a(0)|$ close to $a_0$, see Fig.\ref{decay_figure}. 


The  equation \eqref{mainampeqfree}  gives us the leading-order 
contributions 
to the frequency of the wobbling:
\begin{align}
\label{ampfreq}
\omega = \omega_0 \left[1-\tfrac{1}{2}v^2 + \tfrac{1}{2}
\zeta_R \, |a|^2 + \order{|a|^4}\right],
\end{align}
with $\zeta_R<0$ as in \eqref{Rek1}. 
(Note that $\omega$ is the frequency of oscillation of the 
``full" field $\phi$, not just of the amplitude $a$.)
 The $|a|^2$-term here is a 
nonlinear frequency shift from the linear frequency $\omega_0=\sqrt{3}$;
as time advances, 
this term decays, slowly, to zero.
The $v^2$-term comes from the transverse  Doppler effect. 
We could have obtained this 
term simply by calculating the wobbling  frequency in the rest
frame and then multiplying the result by the relativistic time-dilation 
factor $\sqrt{1-v^2}$ (which becomes $1-\tfrac{1}{2}v^2$ for small
$v$).

 \section{Concluding remarks}
\label{conclusions} 

In this paper, we have formulated a singular perturbation expansion
for the wobbling kink of the $\phi^4$ model. Unlike
the previously published singular perturbation
theories based on the Krylov-Bogoliubov 
and Lindstedt methods, our approach exploits the existence 
of multiple space and time scales in the kink+radiation system.
Some aspects of our scheme are standard to the method 
of multiple scales; some other ones (e.g. the appearance of the
quasisecular terms) are less traditional. We particularly emphasise our
novel treatment of the long-range radiation and the infinite
propagation speed paradox. The final result of the asymptotic analysis 
is the amplitude equation for the wobbling mode, Eq.\eqref{mainampeqfree}. 
Using this equation, we evaluate the nonlinear frequency shift and decay rate
of the wobbler.

The coupling of a spatially localised temporally periodic
excitation to radiation modes via a nonlinearity was 
discussed previously in several contexts. 
In particular, Ref.\cite{PKA}
described the decay of the internal mode of the nonlinear
Schr\"odinger soliton, in the equation
with a general nonlinearity. 
(For rigorous estimates, see e.g. \cite{rigorous}.)
In Ref.\cite{ABP}, the dynamics of the soliton's internal mode
 was considered in the
nonlinear Schr\"odinger 
 equation with the parametric forcing and damping.
Next, the authors of Ref.\cite{SW} studied the  
persistence of a  localised linear impurity mode in the cubic 
Klein-Gordon equation. [We note that although our Eq.\eqref{phi4}
can also be cast in the form of an equation with an impurity
potential --- by letting $\phi=\phi_0 + \chi$ ---
the resulting Klein-Gordon equation satisfied by $\chi$ does not fall into
the class of systems covered by the analysis in that paper.]
We also mention an earlier article \cite{Sigal} where a similar problem
was considered for the nonlinear wave equation.

We conclude our study by producing the perturbation expansion 
of the wobbling kink in terms 
of the original variables:
\begin{multline} \label{phi_sum} 
\phi(x,t) = \tanh \left( \frac{1-3|a|^2}{\sqrt{1-v^2}} \xi \right) \\
+ a \sech \xi \tanh \xi e^{i\omega_0(t-v \xi) } + c.c. \\ 
+ 2 |a|^2\sech^2 \xi \tanh \xi 
+ a^2 f_1(\xi)e^{2i\omega_0(t-v \xi) } + c.c. 
+ \order{|a|^3}.
\end{multline}
Here $\xi=x-vt$;
the complex function $a(t)$ satisfies an ordinary differential equation 
\eqref{mainampeqfree},
  and $f_1(\xi)$ is given by
Eq.\eqref{f1}. Note that we have incorporated two $X_0 \sech^2 X_0$ terms 
of the sum \eqref{varphi2} into the variable width of the kink.
The expansion \eqref{phi_sum} is only valid at the length scale
$|\xi|=\order{1}$;
for larger distances one has to use the outer expansions 
\eqref{outexp} with coefficients determined in section 
\ref{long_scale}.

The first term in \eqref{phi_sum} describes a moving
nonoscillatory kink
with the width decreasing (to the  value of $\sqrt{1-v^2}$)
on the timescale $t \sim |a|^{-2}$. 
The second term  describes the wobbling mode;
the third
gives the 
quasistationary correction to the shape of the kink
induced by the wobbling, and the last term
 accounts for the second-harmonic radiation from 
the wobbler.

The first term in \eqref{phi_sum} is manifestly Lorentz-covariant.
The other terms can also be  cast in the
relativistically-covariant form 
if we replace $\xi$ with $ \xi/ \sqrt{1-v^2}$ in
$\sech \xi$ and $\tanh \xi$
(this is correct to the order of $v^2$), and write 
$ae^{i\omega_0(t-v \xi)}$ as 
\[
|a| \exp \left[
i\omega_0(1+ \tfrac12 \zeta_R
|a|^2) \frac{t-vx}{\sqrt{1-v^2}}
\right].
\]
Here we used Eq.\eqref{ampfreq} and neglected terms of order $|a|^4x$
and $|a|^5t$.
(We remind the reader that $v$ and $|a|$ 
are considered to be small quantities, of the same order of smallness.)

\begin{acknowledgments}
We thank Jakub Lis, Dmitry Pelinovsky, and
Yaroslav Zolotaryuk 
for providing us with useful references.
I.B. was supported by the National Research Foundation of South Africa under
grant 2053723. O.O. was supported by funds provided by the NRF 
and the University of Cape Town.
\end{acknowledgments}

\appendix
\section{Numerical method}
\label{A}

In this Appendix we outline some relevant details
of the numerical method that we used to verify predictions of our
theoretical
analysis. 

Equation \eqref{phi4} was  simulated using an explicit finite-difference scheme
on a grid of step size $\Delta x=0.1$  and $\Delta t=0.05$.
The simulations were performed on the interval $-L<x <L$, where $L$ 
was chosen large enough to prevent the kink exiting the domain of 
integration. (Typical values of $L$ were of the order of $1000$.)
We imposed the free-end boundary conditions.

In order to prevent the radiation reflecting back from the boundaries of the 
system, damping was introduced near 
the edges to absorb the radiation.  That is, 
we added to the $\phi^4$ equation 
an absorbing term ${\tilde \gamma}(x) \phi_t$, with 
\begin{align*}
{\tilde \gamma} (x) &=
\begin{cases}
\left[\frac{x-(L-100)}{100}\right]^4  & \text{for $x \ge L-100$;} \\
\left[\frac{x+(L-100)}{100}\right]^4  & \text{for $x \le -L+100$;} \\
0 & \text{otherwise.}
\end{cases}
\end{align*}

The position $x_0(t)$ of the 
wobbling kink was determined from the location of the
zero crossing. The 
amplitude of the wobbling mode was measured by taking the profile 
$\phi(x,t)$, subtracting the reference kink $\tanh[x-x_0(t)]$,
and assuming the odd component of what remains to be the first-harmonic 
wobbling mode, $a\sech X_0\tanh X_0e^{i\omega_0 \tau}+ c.c.$.  This 
technique, of course, furnishes only a first-order approximation
to the amplitude because of the 
higher order terms in the perturbation expansion.  Interpolation and 
smoothing were applied to counter the effects of the discreteness of 
the $x$ values and the various oscillations occurring on the fast 
time scale.


\begin{thebibliography}{4}

\bibitem{Stats}
A. R. Bishop and T. Schneider (eds).
{\it Solitons and condensed matter physics.\/}
  Springer-Verlag, Berlin, 1978;
  A. R. Bishop. Solitons and Physical Perturbations. In:
{\it Solitons in Action.\/} Editors K. Lonngren and A. Scott.
Academic Press, New York, 1978  
 

\bibitem{QFT_books} R. Rajaraman, {\it Solitons and
Instantons.\/} North-Holland, Amsterdam, 1982;
N. Manton and P. Sutcliffe. {\it Topological Solitons.\/}
Cambridge University Press, Cambridge, England, 2004

\bibitem{cosmology} 
 A. Vilenkin and E. P. S. Shellard. {\it Cosmic Strings and Other
 Topological Defects. \/}
 Cambridge University Press, Cambridge, England, 1994
 

\bibitem{Ferromagnetic_DW}
B. A. Ivanov, V. I. Krasnov and E. V. Tartakovskaya,
Zh. Tekh. Fiz. {\bf 13} 341 (1987);
B. A. Ivanov, A. N. Kichizhiev, and Yu. N. Mitsai,
Sov. Phys. JETP {\bf 75} 329 (1992); 
 B. A. Ivanov, N. E. Kulagin, K. A. Safaryan, Physica B {\bf 202} 193 (1994)
 

\bibitem{Ferroelectric_DW}
 M. A. Collins, A. Blumen, J. F. Currie, and J. Ross,
    Phys. Rev. B {\bf 19} 3630  (1979);
 J. F. Currie, J. A. Krumhansl, A. R. Bishop, S. E. Trullinger,
    Phys. Rev. B {\bf 22} 477  (1980)
   
    
\bibitem{Aubry} S. Aubry, J. Chem. Phys. {\bf 64} 3392 (1976) 

\bibitem{Ferroelectric_exc}
J. A. Krumhansl and J. R. Schrieffer, Phys. Rev. B {\bf 11} 3535 (1975);
G. F. Mazenko and P. S. Sahni, Phys. Rev. B {\bf 18} 6139  (1978);
P. S. Sahni and G. F. Mazenko, Phys. Rev. B {\bf 20} 4674  (1979)

\bibitem{hydrogen} 
Y. Kashimori, T. Kikuchi, K. Nishimoto, J. Chem. Phys. {\bf 77}
1904 (1982);
V. Ya. Antonchenko, A. S. Davydov and A. V. Zolotaryuk,
Phys. Status Solidi (b) {\bf 115} 631 (1983);
E. W. Laedke, K. H. Spatschek, M. Wilkens, Jr., and A. V. Zolotaryuk,
Phys. Rev. A {\bf 32} 1161 (1985);
M. Peyrard, St. Pnevmatikos, N. Flytzanis, Phys. Rev. A {\bf 36} 903
(1987);
J. Halding and P. S. Lomdahl, Phys. Rev. A {\bf 37} 2608 (1988);
E. S. Nylund and G. P. Tsironis, Phys. Rev. Lett. {\bf 66} 1886 (1991);
P. Woafo, R. Takontchoup, A. S. Bokosah, Journ. of Physics and Chemistry of
Solids, {\bf 56}  1277 (1995); 
 A. V. Zolotaryuk, M. Peyrard, K. H. Spatschek,
Phys Rev E {\bf 62} 5706 (2000);
Y. F. Cheng, Chaos, Solitons and Fractals {\bf 21} 835 (2004);
S. Waplak, W. Bednarski and A. Ostrowski, Acta Phys. Polon.
{\bf 108} 261 (2005);
X. F. Pang , Y. P. Feng, H.-w. Zhang, and A. M. Assad,
J. Phys.: Condens. Matter {\bf 18} 9007 (2006);
 Nguetcho AST, Kofane TC, European Phys. Journ.  B {\bf 57}  411 (2007) 

\bibitem{CDW} M. J. Rice, A. R. Bishop, J. A. Krumhansl, and S. E. Trullinger,
Phys. Rev. Lett. {\bf 36}, 432 (1976);
M. J. Rice, Phys Lett A {\bf 71} 152 (1979);
M. J. Rice and J. Timonen, Phys Lett A {\bf 73} 368 (1979) 

\bibitem{RM} 
M. J. Rice and E. J. Mele, Solid State Commun.
 {\bf 35} 487 (1980)
 
\bibitem{phase_transitions} 
W. H. Zurek, Nature {\bf 317} 505 (1985);
J. Dziarmaga,
Phys Rev Lett {\bf 81} 1551 (1998) 


\bibitem{QFT_original} 
R. F. Dashen, B. Hasslacher and A. Neveu, Phys Rev D {\bf 10}, 4139 (1974);
 A. M. Polyakov, JETP Letters {\bf 20} 194 (1974);
 R. Rajaraman and E. J. Weinberg, Phys Rev D {\bf 11} 2950 (1975);
J. Goldstone and R. Jackiw, Phys Rev D {\bf 11} 1486 (1975);
R. Jackiw and C. Rebbi, Phys Rev D {\bf 13} 3398 (1976)

\bibitem{bags} 
W. A. Bardeen, M. S. Chanowitz, S. D. Drell, M. Weinstein
and T. M.-Yan, Phys. Rev. D {\bf 11} 1094 (1975);
    D. K. Campbell and Y.-T. Liao,
Phys. Rev. D {\bf 14} 2093 (1976)
    
    
\bibitem{QFT_recent}
J. S. Rozowsky and C. B. Thorn, Phys Rev Lett {\bf 85} 1614 (2000); 
G. Mussardo, V. Riva, G. Sotkov, Nucl Phys B {\bf 670} 464 (2003);
A. S. Goldhaber, A. Litvintsev, P. van Nieuwenhuizen,
Phys Rev D {\bf 67} 105021 (2003); 
Y. Bergner and L. M. Bettencourt, Phys Rev D {\bf 69} 045002 (2004);
M. Salle, Phys Rev D {\bf 69} 025005 (2004);
D. Chakrabarti, A. Harindranath, J. P. Vary, 
Phys Rev D {\bf 71} 125012 (2005);
S. Dutta, D. A. Steer, and T. Vachaspati,
Phys. Rev. Lett. {\bf 101} 121601 (2008)

\bibitem{collisions}
A. E. Kudryavtsev, JETP Lett. {\bf 22} 82 (1975);
M. J. Ablowitz, M. D. Kruskal, J. F. Ladik, SIAM J. Appl. Math. {\bf 36} 428
(1979);
T. Sugiyama, Prog. Theor. Phys. {\bf 61} 1550 (1979);
M. Moshir, Nucl. Phys. B {\bf 185} 318 (1981);
C. A. Wingate, SIAM J. Appl. Math. {\bf 43} 120 (1983);
R. Klein, W. Hasenfratz, N. Theodorakopoulos, W. Wunderlich, Ferroelectrics
{\bf 26} 721 (1980);
D. K. Campbell, J. F. Schonfeld, and C. A. Wingate, 
Physica D {\bf 9} 1 (1983);
T. I. Belova, A. E. Kudryavtsev, Physica D {\bf 32} 18 (1988);
P. Anninos, S. Oliveira, R. A. Matzner, Phys. Rev. D {\bf 44} 1147 (1991);
T. I. Belova, Physics of Atomic Nuclei {\bf 58} 124 (1995);
T. I. Belova, A. E. Kudryavtsev,
Uspekhi Fizicheskikh Nauk,  {\bf 167}   377 (1997)

\bibitem{Getmanov}
B. S. Getmanov, JETP Lett. {\bf 24} 291 (1976)

\bibitem{Belova} T. I. Belova, ZhETF {\bf 109} 1090 (1996)

\bibitem{Rice} M. J. Rice, Phys Rev B {\bf 28} 3587 (1983)

\bibitem{Segur}
H. Segur, J. Math. Phys. \textbf{24}, 1439 (1983)

\bibitem{Malomed} B. A. Malomed, J. Phys. A: Math. Gen. {\bf 25}  755
(1992)

\bibitem{Manton}
N. S. Manton and H. Merabet, Nonlinearity {\bf 10} 3 (1997)

\bibitem{PelSlu}
R. Pe{\l}ka, Acta Phys. Polon. {\bf 28} 1981 (1997);
M. \'Slusarczyk, Acta Phys. Polon. {\bf 31} 617 (2000)

\bibitem{Roman1} 
T. Roma\'nczukiewicz, Acta Phys. Polon. {\bf 35} 523 (2004)


\bibitem{radiation} 
T. Roma\'nczukiewicz, Acta Phys. Polon. {\bf 36} 3877 (2005);
T. Roma\'nczukiewicz, J. Phys. A: Math. Gen. {\bf 39} 3479 (2006);
P. Forg\'acs, \'A. Luk\'acs, and T. Roma\'nczukiewicz, Phys. Rev. D 
{\bf 77} 125012 (2008)




\bibitem{Nayfeh}
A. H. Nayfeh, {\it Introduction to Perturbation Techniques.\/}
Wiley-Interscience, New York, 1993


\bibitem{Kiselev_free}
O. M. Kiselev, Russian J. Math. Phys. {\bf 5} 29 (1997)


\bibitem{Kiselev_perturbed}
O. M. Kiselev, Siberian Mathematical Journal {\bf 41} 2 (2000)


\bibitem{Sukstanskii} A. L. Sukstanskii and K. I. Primak,
Phys Rev Lett {\bf 75} 3029 (1995)



\bibitem{PKA} D. E. Pelinovsky, Y. S. Kivshar, V. A. Afanasjev,
Physica D {\bf 116} 121 (1998)

\bibitem{rigorous}
V. S. Buslaev, C. Sulem, Ann. I. H. Poincar\'e AN{\bf 20} 419 (2003);
S. Cuccagna, J. Differential Equations {\bf 245} 653 (2008);
S. Cuccagna, Physica D {\bf 238} 38 (2009)

\bibitem{ABP}  N. V. Alexeeva, I. V. Barashenkov and D. E. Pelinovsky,
 Nonlinearity 1999, {\bf 12}, 103-140
 

 \bibitem{SW}
 A. Soffer, M. I. Weinstein, Invent. math. {\bf 136} 9
(1999)

\bibitem{Sigal} I. M. Sigal, Commun. Math. Phys. {\bf 153} 297 (1993)

\bibitem{OB}
O F Oxtoby and I V Barashenkov,
Resonantly driven wobbling kinks. 
(The next arXiv submission.) 


\end{thebibliography}
\end{document}